\documentclass[%
 reprint,
 amsmath,amssymb,
 aps,
]{revtex4-2}

\usepackage{graphicx}
\usepackage{dcolumn}
\usepackage{bm}

\begin{document}

\preprint{APS/123-QED}

\title{Magnetotransport in a perturbed periodic antidot superlattice}

\author{Atahualpa S.~Kraemer}
\email{ata.kraemer@gmail.com}
\affiliation{Physics Departament, Faculty of Science,
Universidad Nacional Aut\'onoma de M\'exico,
Ciudad Universitaria, M\'exico D.F. 04510, Mexico
}

\author{Alan Rodrigo Mendoza Sosa}
\affiliation{Physics Departament, Faculty of Science,
Universidad Nacional Aut\'onoma de M\'exico,
Ciudad Universitaria, M\'exico D.F. 04510, Mexico
}

\date{\today}

\begin{abstract}

We study a 2-dimensional model for an antidot periodic superlattice with perturbed positions of the antidots. To do so we use a quasiperiodic LG model obtained from a 3-dimensional billiard model. Our results show that infinite drifting trajectories present in the periodic antidot models disappear, but if the perturbation is small enough, those trajectories remain for long times. The probability to visit the region of the phase space where electrons have ballistic behavior tends to $0$ as the length of the drifting trajectories tends to infinity, leading to separation of these regions in phase space. As a result, we infer that the particles follow Levy walks and the system has superdiffusive behavior for short times. The superdiffusive exponent is correlated to the length of time where the superdiffusive behavior is present.    
\end{abstract}

\maketitle

Antidot superlattices are artificial periodic nanostructures of obstacles (antidots) where the potential is higher than in the immediately surrounding region, so that a two-dimensional electron gas is repelled. They were first fabricated by ion-beam implantation~\cite{ensslin1990magnetotransport} and then by etching holes in a periodic array into a conductive material~\cite{Weiss1991}. Since then, there have been many experiments to produce and study the properties of such heterostructures~\cite{Noeckel1997, Gmachl1998, Lacey2001, Fukushima2004, Andreasen2009, Harayama2011, Cao2015, maier2017ballistic}. It is expected that they can be treated with a classical approach if their scale is larger than the Fermi wavelength of electrons~\cite{FlieextbackslashSer1996, Boggild1998, Christensson1998, Kuzmany1998, Mendoza2008}. In addition, at low temperature electrons 
can be considered as non-interacting~\cite{Schuster1997, Kouwenhoven1997, Buks1998, Kuzmany1998, Avinun-Kalish2005}. Thus, a billiard model represent these materials well~\cite{Lorke1991, Weiss1991}.

In the last decades, billiard models have shown to be good models of optoelectronic devices~\cite{Noeckel1997, Gmachl1998, Lacey2001, Fukushima2004, Andreasen2009, Harayama2011, Cao2015}, predicting useful chaotic properties. On the other hand, they have been studied for decades by physicists and mathematicians, but only relatively recent by experimental realizations. They were performed using acoustic and optical waves in cavities~\cite{Stockmann1990, Richter1999, Schaadt1999, Schaadt2003, Dietz2015, Lawniczak2017}. One of the most studied billard is the Lorentz gas (LG), which was first suggested as a model of electrons in a metal~\cite{lorentz1905motion}. Since then there has been a great effort to understand its diffusive properties~\cite{dettmann2014diffusion}, but only now we can observe experimentally these properties using antidot superlattices. 

\begin{figure}
    \centering
    \includegraphics[width=220pt]{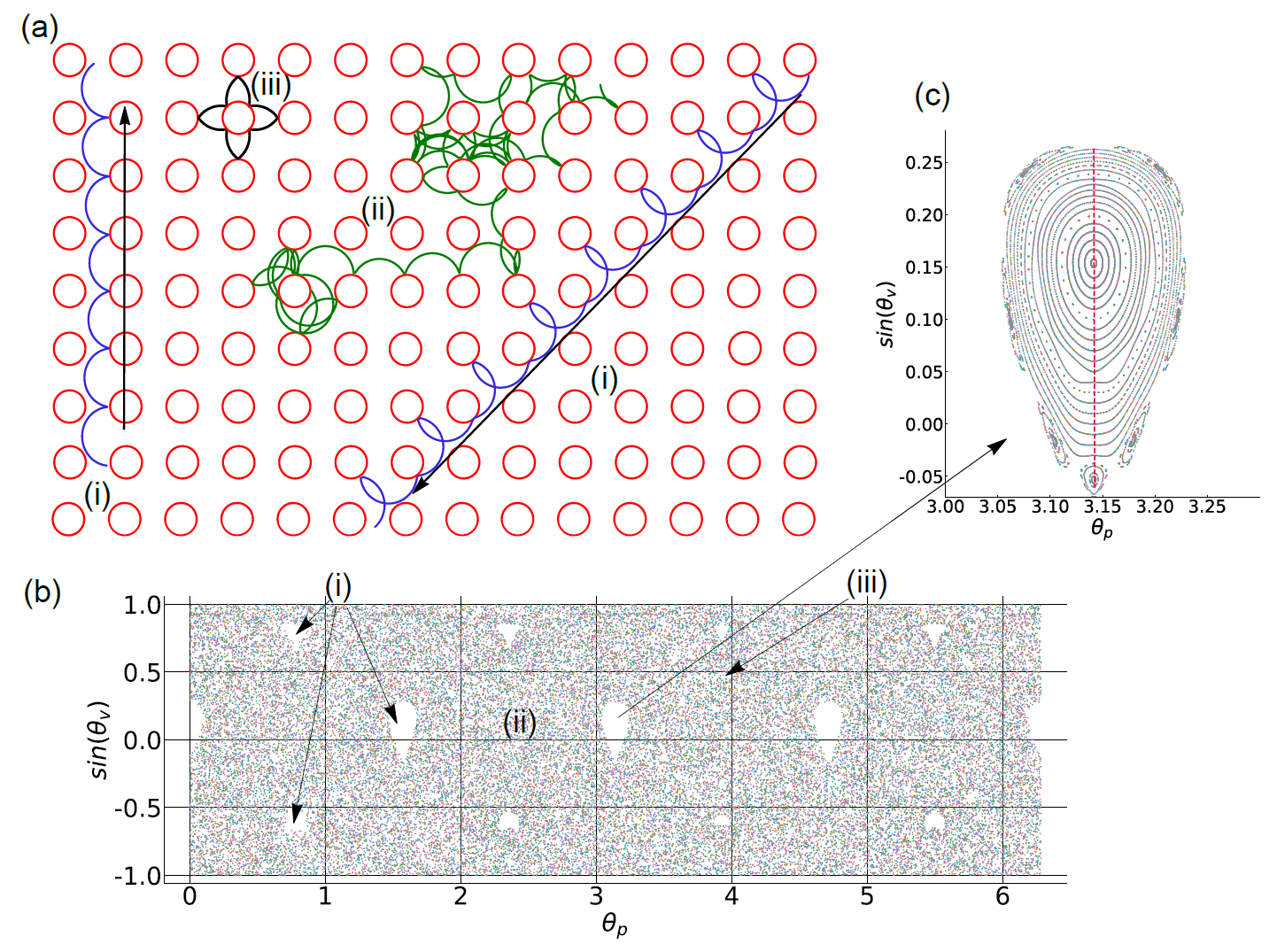}
    \caption{(a) Antidot superlattice with some trajectories. (b)  Poincaré Map of the super lattice. The islands correspond to the trajectories marked with label (i), the chaotic trajectories are labeled with (ii) and flowers are marked with label (iii). (c)  Poincaré Map of some trajectories inside the island.}
    \label{fig:Poincare-trajectories}
\end{figure}

For these reasons, LGs with a magnetic field have been largely studied~\cite{Bobylev1995, FlieextbackslashSer1996, Kuzmany1998, Baskin1998, Bobylev2001, Dmitriev2001, Cheianov2003, Schirmacher2015}. Recently, it has been investigated the transition to localization when the magnetic field is increased in a random LG~\cite{Fogler1997, baskin1978, Kuzmany1998, Schirmacher2015}. In the localization limit there is sub-diffusion similarly to random LG with linear trajectories when density is decreased, but the subdiffusion exponent is different.
On the other hand, in the periodic LG, it has been shown~\cite{fleischmann1992magnetoresistance, FlieextbackslashSer1996} that certain trajectories move ballistically (see figure \ref{fig:Poincare-trajectories} (a)). These trajectories are stable even when the obstacles potential is perturbed~\cite{datseris2019robustness}.

So, on one hand, magnetotransport in disordered systems can be subdiffusive, and on the other hand, for periodic systems it is possible to find ballistic motion. 
A natural question is to ask what occurs when the position of antidots is perturbed.  

In the case of LGs without a magnetic field, the trajectories that hit an obstacle are dense~\cite{sinai1979ergodic}, thus, any initial volume of the phase spaces will fill ergodically the whole phase space. A way to visualize this is by a Poincaré map. The  Poincaré Map of a periodic LG can be obtained by considering each successive collision with an obstacle. Then the important variables are the position of a collision on the obstacle $\theta_p$ and the velocity at the time of hitting the obstacle. Since the speed of the particles in a LG is constant, the velocity is determined by the angle between the velocity vector and the tangent of the obstacle at the point of collision, $\theta_v$. Then, to make a  Poincaré Map of a trajectory, we measure $\theta_p$ and $\sin(\theta_v)$ for each collision in a long trajectory and we plot those points. The  Poincaré Map will be embedded in a rectangle of area $2\pi \times 1$. 

For periodic LG with a magnetic field, the behavior of the  Poincaré Map depends on the radius of the trajectories and the density of the obstacles. We obtain 3 main kinds of trajectories (see figure \ref{fig:Poincare-trajectories}). Chaotic trajectories fill densely the majority of the rectangle in the  Poincaré Map, except for some relatively small areas that we call ``islands''. Those trajectories are diffusive. Trajectories inside the islands follow drifting trajectories lead to ballistic motion.
A trajectory that begins inside the island will not fill the island densely as figure \ref{fig:Poincare-trajectories}(c) shows. It is not possible to go from a chaotic trajectory to a drifting trajectory and vice-versa. Flowers are localized trajectories that do not contribute to the diffusion. Those trajectories are unstable when the magnetic field is low. 

\begin{figure}
    \centering
    \includegraphics[width=220pt]{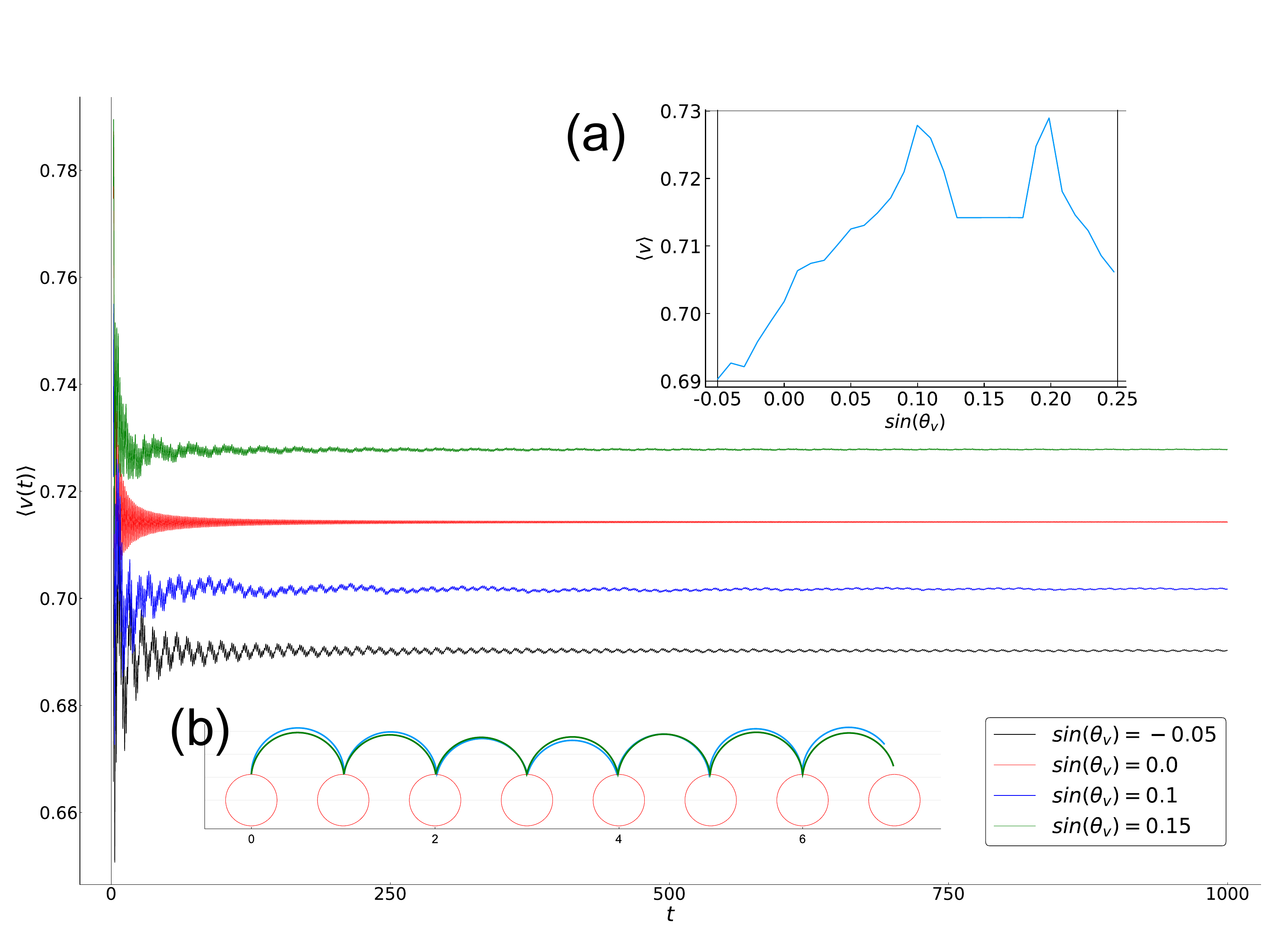}
    \caption{(a) Average velocities for different drifting trajectories. (b) Two trajectories considered}
    \label{fig:velocity-driffting-trajectories}
\end{figure}

Trajectories inside an island have all the same drifting direction. Thus, if we slightly perturb the initial conditions of such particles, we will still have drifting trajectories in the same direction. However, the mean drifting velocity, defined as $V_d(t) = \lim_{t \rightarrow \infty} \frac{x(t) -x(0)}{t}$, is a function of the initial conditions of the particle. In figure \ref{fig:velocity-driffting-trajectories} we show different average velocities of the same drifting direction, corresponding to the island plotted in figure \ref{fig:Poincare-trajectories}(c).  In the inset of figure \ref{fig:velocity-driffting-trajectories} we plot the average velocity for the initial position on the obstacle equals to $\pi$ (shown in figure \ref{fig:Poincare-trajectories} (c) as a dashed line), and the initial velocity varying between $-0.05$ and $0.25$, which are $\theta_v$ values within the island of the figure \ref{fig:Poincare-trajectories}(c). There are two maxima for the drifting velocities. Note that they correspond to the same trajectory, but they do not correspond to the periodic trajectory which is close to $\sin(\theta) = 0.15$. Trajectories close to the periodic trajectory have almost the same drifting velocity. It is an open question to explain this behavior.

\paragraph{Model}
\label{model}

Based on the idea that perturbation on the initial conditions do not affect considerable the drifting trajectories, we could expect that small perturbations on the position of the obstacles would not affect the drifting trajectories either, and so, the mean square displacement would be ballistic, which means $\langle x^2(t) \rangle \sim t^2$. However, super-diffusion appears because particles collide always on similar values of $\theta_p$ and with the same range of velocities, then, we the obstacles in a particular position. It is not clear if any perturbation on the position of obstacles would affect the hitting position, destroying in long therm drifting trajectories, or there is a minimum value of perturbation to distroy those trajectories.  

A periodic array of obstacles can be seen as a periodic potential $V(\vec{x})$. A perturbation on this periodic function can obtain by adding a small random vector of length $\delta$ to their positions. In such a case, instead of having $|V( \vec{x} + \tau ) - V( \vec{x})| = 0$, we would have $|V(\vec{x}+ \tau )-V(\vec{x})| < \delta$. In this way, a minimal perturbation on the potential function should produce a function that for every $\delta >0$ exist a $\tau$, such that $|V(\vec{x}+ \tau ) - V(\vec{x})|<\delta$ for every $\vec{x}$. This is an almost-periodic function. A quasiperiodic array of obstacles are also almost-periodic, so, we can think of quasiperiodic LGs as perturbed periodic LGs. Studying quasiperiodic systems have two advantages over random environments: (i) The size of the system needed to study quasiperiodic LGs is finite, since we can embed the quasiperiodic array in to a higher dimension periodic system~\cite{PRL-Ata-David, kraemer2015horizons}, while for random environments we need in principle an infinite system. (ii) The uniform distribution for the initial conditions is not well defined in random environments and neither is  Poincaré Map, while in the quasiperiodic case we can use a higher dimensional periodic billiard and define homogeneous distribution in the periodic cell. 

\begin{figure}
    \centering
    \includegraphics[width=150pt]{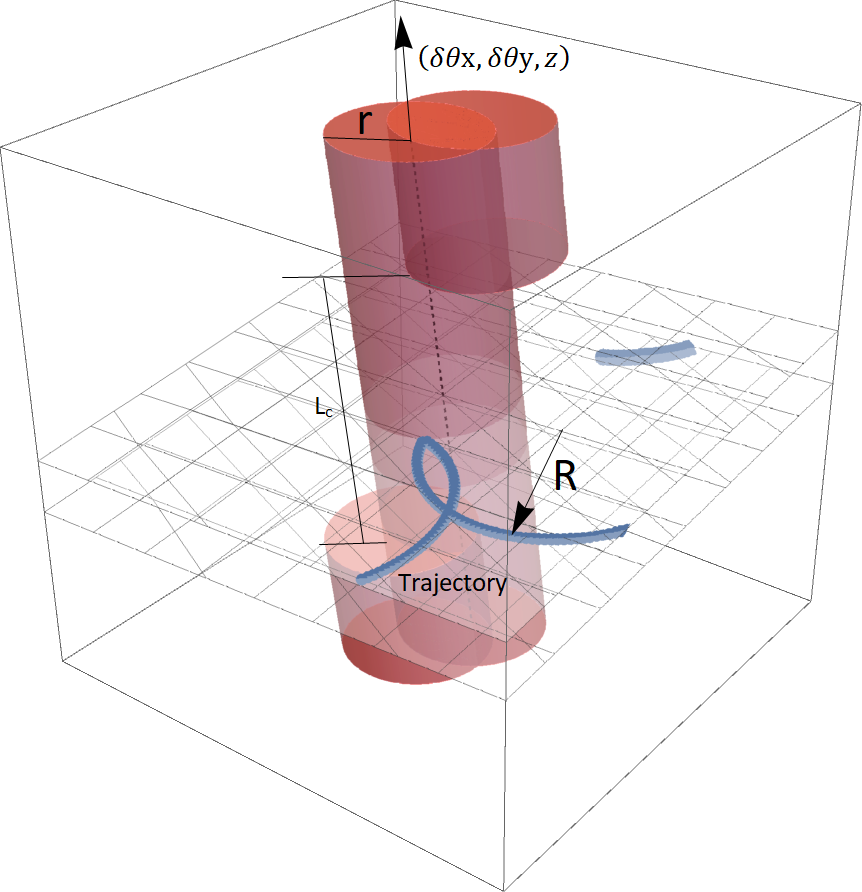}
    \caption{Billiard used to simulate a quasiperiodic LG with circular motion of the particles.}
    \label{fig:model}
\end{figure}

The higher-dimensional billiard model equivalent to the LG in its simplest form consists of a 3-dimensional cylinder with hardcore potential embedded in a cube with periodic boundary conditions. This cylinder is generically longer than the distance between two parallel faces of the cube in the direction of the axis of the cylinder, so that part of the cylinder is outside of the cube. It is then necessary to apply periodic boundary conditions on the cylinder forming a set of at least 3 cylinders. If the radius of the cylinder is less than a critical value, then there will be exactly 3 cylinders after applying periodic boundary conditions, one that passes through the center of the cube and two smaller ones, whose axes cross each one of the faces of the cube (see figure \ref{fig:model}). This critical value depends on the angle of inclination of the cylinder. If the cylinder is vertical (in such case the system is periodic), then the critical value is $\frac{1}{2}L$, where $L$ is the length of the cube. This critical value is reduced when the inclination of the cylinder increases. 

If the axis of the cylinder is completely rational, then the system will be a periodic LG, while a completely irrational inclination will produce a quasiperiodic array of obstacles. Finally, to perform the simulations, the particles must be constrained to move on planes perpendicular to the cylinder. They will exhibit specular reflections when colliding with the cylinder, otherwise they will move following circular trajectories with a radius of trajectory $R = \frac{mv}{qB}$, where $m$ and $q$ are the mass and charge of the electron respectively, $v$ is the velocity of particles, and $B$ is the magnetic field. 

In this model, if the cylinder is vertical, the equivalent LG model will be a periodic LG as shown in figure \ref{fig:Poincare-trajectories} (a). 
To produce a perturbation on obstacle positions we slightly incline the cylinder. In this way, we obtain a quasiperiodic array of obstacles. We notice that in order to produce a billiard equivalent to a quasiperiodic LG, the inclination (perturbation) of the cylinder must be completely irrational.  

The parameters of this model are the direction (inclination) of the cylinder axis $\vec{x}_c$, the speed of the particles $|v|$, the radius $r$ of the cylinder, the length $L$ of the cube, and the radius $R$ of trajectories (proportional to $1/B$). Without loss of generality we can set $|v| = 1$ and $L = 1$ because the system is re-scalable. We also set the radius of trajectories at $R = 0.506$, and the radius of cylinder at $r = 0.28$. For these values there are islands in the Poincaré map for the periodic LG of considerable area. To perform the simulation we used $5\times 10^3$ particles during $10^5$ units of time. Finally, the direction of the cylinder was first $\vec{x}_c = (0,0,1)$ for the periodic case, and then we perturbed the system using $\vec{x} = (\delta \theta \cdot \sqrt{2}, \delta \theta \cdot \pi, 1)$, where $\delta \theta $ was the perturbation. We use this vector as the direction of the cylinder axis since it is completely irrational for most values of $\delta \theta$.   

\begin{figure}
    \centering
    \includegraphics[width=220pt]{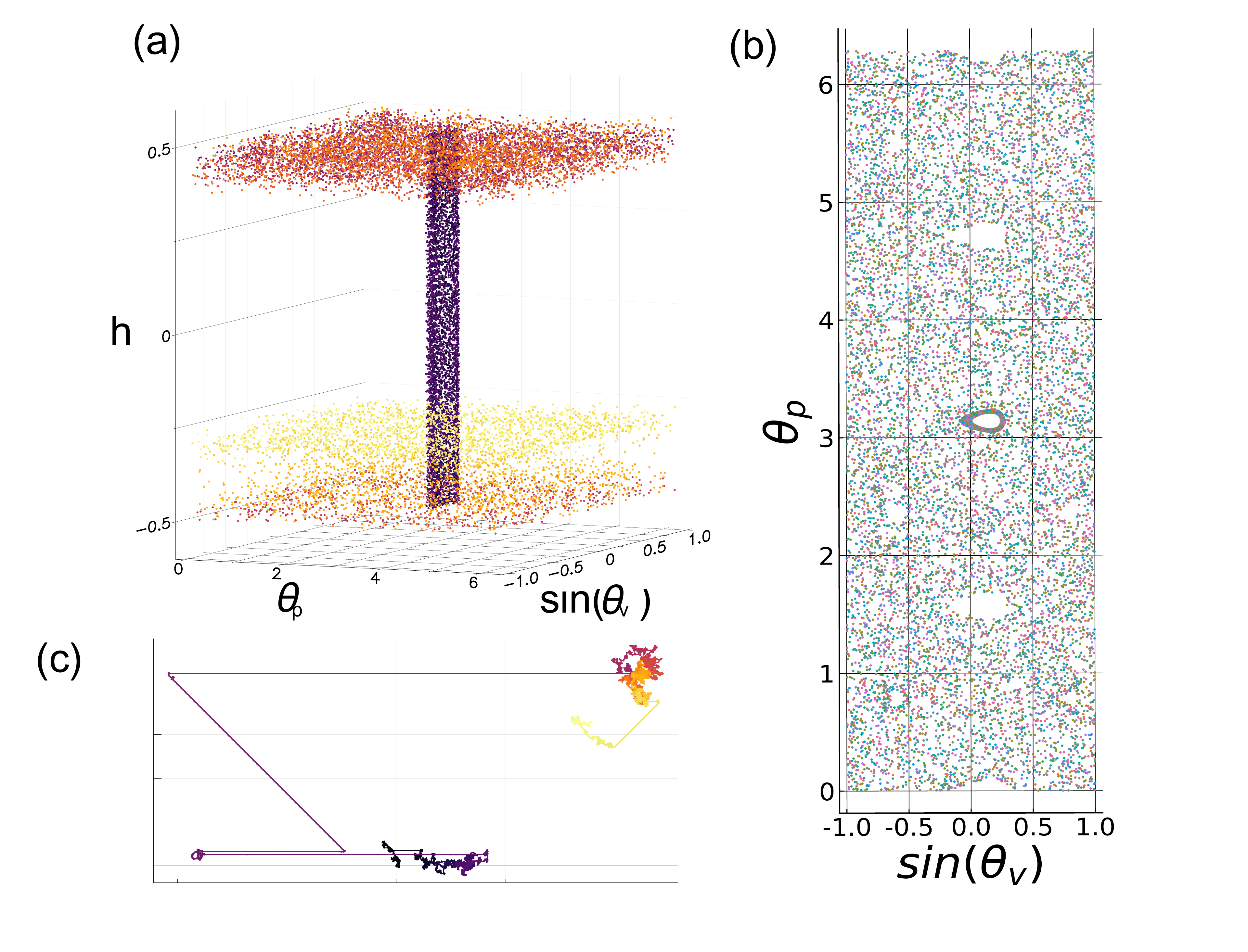}
    \caption{Poincaré map of a long trajectory in a perturbed system. (a) 3D  Poincaré Map of a quasiperiodic LG.  (b) 2D projection of the  Poincaré Map. (c) Typical trajectory of in the perturbed superlattice.}
    \label{fig:Poincare-quasi}
\end{figure}

The  Poincaré Map of a quasiperiodic LG is no longer a 2-dimensional plot of finite area. Because the real model is a 3D billiard. The extra dimension is given by the height $h$ on the cylinder where the particle collides. Then, the  Poincaré Map is obtained by a plot of $\theta_p$, $sin(\theta_v)$ and $h$. If the cylinder is vertical, then the 3-dimensional Poincaré Map will be only one slice of the 3D map. If we perturb the system by making a small inclination of the cylinder, we do not expect a big difference with respect to the periodic case, so we may study the 2D projection of the  Poincaré Map. Figure \ref{fig:Poincare-quasi} shows the 2 and 3-dimensional version of the  Poincaré Map when a small perturbation ($\delta \theta = 0.0001$) is applied.   

As we can see in the 2-dimensional  Poincaré Map (figure \ref{fig:Poincare-quasi} (b)), it is very similar to the periodic case, except that one trajectory seems like a mixture of chaotic and drifting trajectories. The islands are no longer separated from the chaotic region, making all the phase space accessible for almost any initial condition. As expected, for short trajectories particles belonging to the chaotic region moves as a random walk visiting almost all the  Poincaré Map except for some ``islands'', and particles inside the islands move following drifting trajectories. However, if trajectories are long enough, then particles can visit both, the chaotic region and the islands. 

In the 3D  Poincaré Map (figure \ref{fig:Poincare-quasi}(a)) it is shown that drifting trajectories move fast along the $z$-axis, while chaotic trajectories move fast in the $x-y$ plane. We can also see that most of the chaotic part is for extrem values of $h$. This is because the transition from drifting to chaotic trajectories only occurs on the planes where two cylinders coexist. Once a particle is in the chaotic region, it can travel in the Poincaré Map to middle values of $h$, so that the  Poincaré Map is filled.

If we perturb the periodic LG, for long trajectories the drifting trajectories vanish falling into chaotic trajectories, so we no longer expect a ballistic motion of electrons. However, for short times drifting trajectories still, exist. 

We observed in the 3D Poincaré Map, drifting trajectories move fast in the $z$ axis, and once they rich an extreme value of $h$, they change to a chaotic trajectory. This means that there is a maximum length for the drifting trajectories. This length can be calculated by the velocity at which the drifting trajectories move along the $z$ axis of the  Poincaré Map, and the length of the central cylinder that does not coexist with the other cylinders $L_f$. This length can be calculated using the length of the cylinder before applying periodic boundary conditions $L_c$. The length of the small cylinders will be $\sim (L_c -L)/2$, so the length we search will be $L_f= L-2 (L_c -L)/2 = 2L-L_c$. Each time a particle crosses a face of the cube in the billiard model, the $z$ position in the  Poincaré Map changes proportional to the slope $m$ of the plane where the particle moves in the direction of the crossed face. The speed at which the particle crosses one face of the cube is proportional and close to the drifting velocity $v_d/2$. Then, the speed at which drifting particles moves along the $z$ axis, is approximately $m v_d/2$, and the maximum time that a drifting trajectory survives will be $t_{max} \sim 2(2L-L_c)/(m_{min} v_d)$, where $m_{min}$ is the minimum of the slopes of the plane in the direction of each of the faces of the cube. 

Thus, we expect a maximum length for the drifting trajectories, depending on the perturbation $\delta \theta$. If the time scale is much longer than the maximum length, we expect normal diffusion. 

On the other hand, for the periodic case, the mean velocity of particles can be well approximated by the velocity of the drifting trajectories, multiplied by the portion of the volume of the phase space where particles follow such drifting trajectories. If we perturb the system, the volume of the phase space where particles follow a drifting trajectory longer than a time $t$ depends on $t$. Call $f(t)$ this portion of volume. Then, the average velocity at which of particles spread will be $v(t) = f(t)*v_d$, where $v_d$ is the drifting velocity in the non-perturbed system. In such a case, we can approximate the mean square displacement as:

\begin{equation}
    \langle x^2(t) \rangle \sim v_d^2 f^2(t) t^2.
    \label{eq = super-diffusion}
\end{equation}

If $f(t)\sim t^{-\alpha/2}$, with $1>\alpha >0$, we will obtain a superdiffusive behavior. We know that $f(x)$ is a monotonically decreasing function, so we can approximate it by a power law: 

\begin{equation}
    \langle x^2(t) \rangle \sim v_d^2 t^{2-\alpha}.
    \label{eq = super-diffusion2}
\end{equation}

\begin{figure}
    \centering
    \includegraphics[width=220pt]{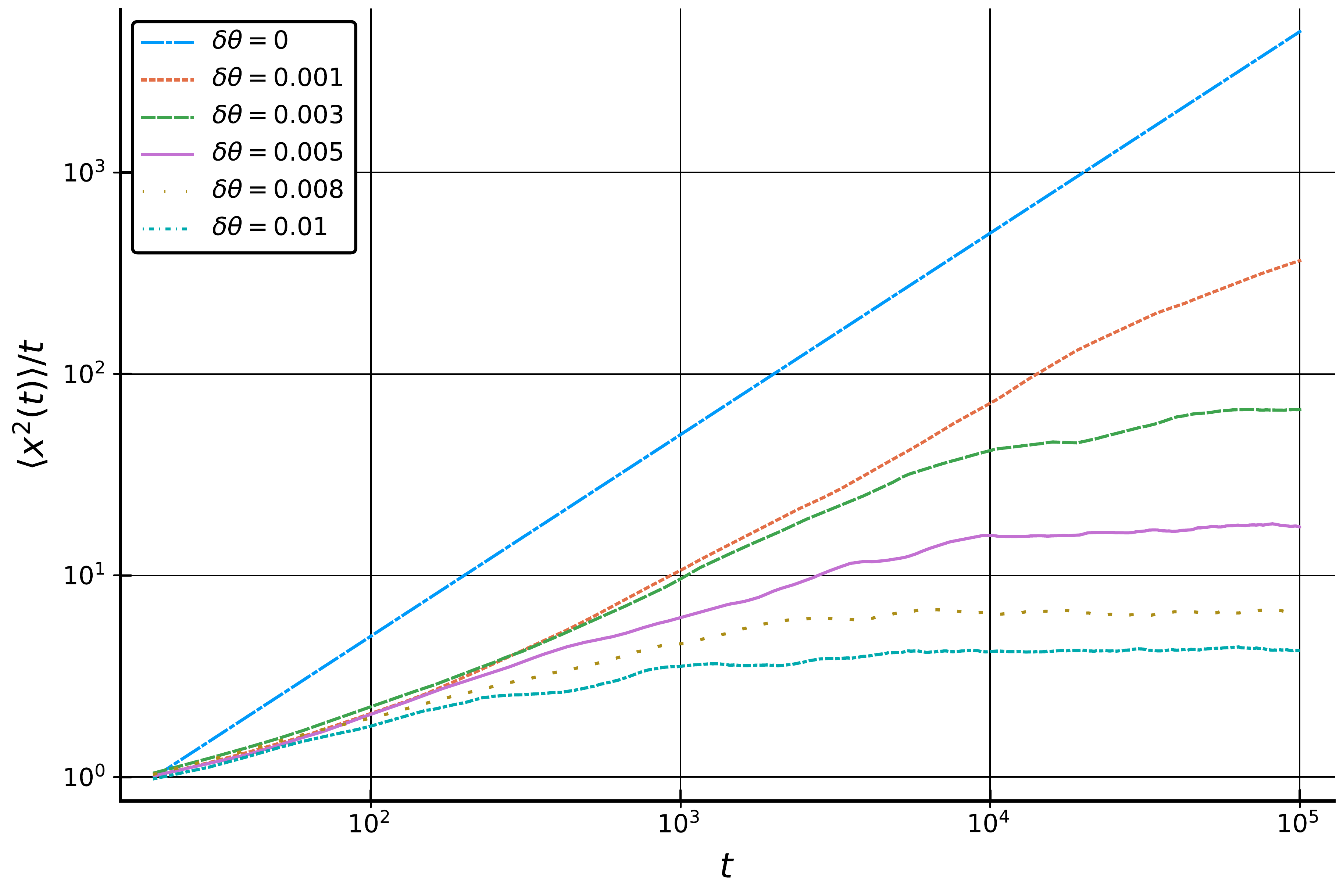}
    \caption{Mean square displacement over time as a function of time for several perturbation values}
    \label{fig:MSD}
\end{figure}

We still must justify $\alpha<1$. We have performed numerical simulations, and we observed that the exponent is always  $0<\alpha<1$. Figure \ref{fig:MSD} shows the evolution of the mean square displacement as a function of time. We divided $\langle x^2(t) \rangle$ by $t$, so $\langle x^2(t) \rangle /t \sim v_d^2 t^{1-\alpha}$, then if $\alpha>1$ the plot will be a decreasing function, while for $\alpha <1$ will be an increasing function. We observe that the length of time where the system is superdiffusive increases when the perturbation is reduced. We can also see that $\alpha$ depends on $\delta \theta$, increasing when $\delta \theta$ decreases.  

In sumary, we analyzed the Poincaré Map and diffusive properties of a perturbed antidot superlattices model with a quasiperiodic model, obtaining superdiffusion for short times due to Levy walks in contrast to the ballistic motion of electrons found in the periodic model. The length of time over which the superdiffusion occurs depends on the perturbation, but also on the drifting velocities. In future work, we will study the effect of variation in the magnetic field on the diffusion of particles in this quasiperiodic model.

\begin{acknowledgments}
We thank David P. Sanders for useful discussions. Financial support is acknowledged from DGAPA-UNAM PAPIIT grant IA106618. ARMS received a fellowship from the DGAPA-UNAM PAPIIT grant IN117117. 

\end{acknowledgments}

%


\end{document}